\documentclass[a4paper,11pt]{article}
\pdfoutput=1 

\usepackage{jcappub} 
\usepackage[T1]{fontenc} 

\usepackage{color}
\usepackage{amsmath,amssymb}
\usepackage{graphicx}
\usepackage{epsfig}
\usepackage{subfigure}

\def\gsim{\ \rlap{\raise 3pt \hbox{$>$}}{\lower 3pt \hbox{$\sim$}}\ }
\def\lsim{\ \rlap{\raise 3pt \hbox{$<$}}{\lower 3pt \hbox{$\sim$}}\ }

\newcommand{\be}{\begin{equation}}
\newcommand{\ee}{\end{equation}}
\newcommand{\bea}{\begin{eqnarray}}
\newcommand{\eea}{\end{eqnarray}}
\def\zb{\mathbb{Z}}
\newcommand{\eqn}[1]{eq.~(\ref{#1})}

\newcommand{\Eqn}[1]{Eq.~(\ref{#1})}

\renewcommand{\thefigure}{\Roman{figure}}

\title{A cosmological pathway to testable leptogenesis}

\author{Bhaskar Dutta$^{a}$,}
\author{Chee Sheng Fong$^{b,b*}$,}
\author{Esteban Jimenez$^{a}$,}
\author{Enrico Nardi$^{c}$}


\affiliation[a]{\it Mitchell Institute for Fundamental Physics and Astronomy, Department of Physics and Astronomy, Texas A\&M University, College Station, TX 77843, USA}

\affiliation[b]{\it Instituto de F\'isica, Universidade de S\~ao Paulo, S\~ao Paulo, Brazil}

\affiliation[b*]{\it Departamento de F\'isica, Pontif\'icia Universidade Cat\'olica do Rio de Janeiro, Rio de Janeiro, Brazil}              
                    
\affiliation[c]{\it  INFN, Laboratori Nazionali di Frascati, C.P. 13, 100044 Frascati, Italy.}

\emailAdd{dutta@physics.tamu.edu}
\emailAdd{cheesheng.fong@gmail.com}
\emailAdd{este1985@physics.tamu.edu}
\emailAdd{enrico.nardi@lnf.infn.it}

\abstract{Leptogenesis could have occurred at temperatures much
  lower than generally thought, if the cosmological history of the
  Universe underwent a period of accelerated expansion, as is
  predicted for example in a class of scalar-tensor theories of
  gravitation. We discuss how non-standard cosmologies can open new
  pathways for low scale leptogenesis. Within these scenarios
  direct tests of leptogenesis could also provide informations on the
   very early times Universe evolution, corresponding to temperatures
   larger than the TeV.}

\begin{document}

\maketitle

\section{Introduction}
\label{introduction}  
The cosmological baryon asymmetry is very elegantly explained via the
leptogenesis mechanism~\cite{Fukugita:1986hr} according to which an
initial asymmetry is generated in lepton number and then partly
converted in a baryon number asymmetry by $B+L$ violating sphaleron
processes~\cite{Kuzmin:1985mm,Rubakov:1996vz} which, above the
temperature of the electroweak (EW) phase transition, proceed with
in-equilibrium rates (for reviews on the leptogenesis mechanism
see~\cite{Davidson:2008bu,Fong:2013wr}).  A very attractive feature of
the standard leptogenesis realization based on the type-I
seesaw~\cite{Minkowski:1977sc,GellMann:1980vs,Yanagida:1979as,%
  Mohapatra:1980yp} is that it provides a semi-quantitative relation
connecting the out-of-equilibrium condition~\cite{Sakharov:1967dj} for
the decays of the heavy right handed (RH) neutrinos with the light
neutrino mass scale. RH neutrino decays can be sufficiently out of
equilibrium if $m_\nu \sim 10^{-2\pm 1}\,$eV, which is in beautiful
agreement with neutrino oscillation data.  On the other hand, type-I
seesaw leptogenesis has also an unpleasant facet.  A lepton asymmetry
is preferably generated in the decay of the lightest RH neutrino
$N_{1}$ since they generally occur at temperatures when the dynamics
of the heavier $N_{2,3}$ neutrino is no more efficient. However,
the CP asymmetry in $N_1$ decays is bounded by the following
relation~\cite{Davidson:2002qv}:
\begin{equation}
  \label{eq:DIbound}
 \left|\epsilon_1\right| \leq \frac{3}{16\pi}\frac{M_1}{v^2}
\frac{\Delta m^2_\oplus}{m_1+m_3}  \,, 
\end{equation}
where $M_1$ is the $N_1$ mass, $v\sim 174\,$GeV is the Standard Model (SM) 
EW breaking vacuum expectation value (VEV),
$\Delta m^2_\oplus\sim 2.4 \times 10^{-3}\,{\rm eV}^2$ is the
atmospheric neutrino mass square difference, and $m_{1,3}$ are the
lightest and heaviest light neutrino masses, which are bounded by
cosmological data to lie not much above $\sim 10^{-1}\,$eV. Since a
minimum CP asymmetry
$\left|\epsilon_1\right| \gsim {\rm few}\times 10^{-7}$ is required to
account quantitatively for the observed baryon asymmetry, the $N_1$
mass cannot lie much below $ 10^9\,$GeV.\footnote{The bound
  \eqn{eq:DIbound} can be somewhat weakened if the RH neutrinos masses
  are not sufficiently hierarchical~\cite{Hambye:2005tk}, if $N_{2,3}$
  decays also contribute to the generation of a lepton
  asymmetry~\cite{Engelhard:2006yg}, or if flavor
  effects~\cite{Abada:2006fw,Nardi:2006fx,Abada:2006ea} play a
  relevant role~\cite{Blanchet:2006be,Racker:2012vw}.  However, the main conclusion
  regarding non-testability of the type-I seesaw leptogenesis model
  does not change.}  The conclusion that the CP asymmetry is too small
to explain the baryon asymmetry if the leptogenesis scale is too low,
implies that direct tests of the standard type-I seesaw leptogenesis
are out of experimental reach. Since the argument does not involve any
cosmological input, it holds regardless of the assumed cosmological
model.

In more generic realization of leptogenesis~\eqn{eq:DIbound} does not
necessarily hold: the simple relation between the CP asymmetries and
the light neutrino masses is in fact quite specific of the type-I
seesaw and is often lost in other models.  
The most direct way to relax this bound is to rescale $v$ in \eqn{eq:DIbound} and this can be realized in model where neutrinos only couple through a neutrinophilic Higgs which obtains a VEV $v_\nu \ll v$ ~\cite{Haba:2011ra,Clarke:2015hta}.
Other examples are the inert scalar doublet model~\cite{Ma:2006km} complemented with heavy Majorana neutrinos~\cite{Racker:2013lua},
as well as many other models, see~\cite{Sierra:2014tqa,
Sierra:2013kba,Racker:2012vw,Hambye:2012fh,AristizabalSierra:2009bh,
GonzalezGarcia:2009qd,AristizabalSierra:2007ur} for a sample list. 
Still, the vast majority of models that attempt to generate the baryon
asymmetry from heavy particle decays are subject to an additional
constraint which, although less tight than the one implied by
\eqn{eq:DIbound}, is much more general.  This constraint stems from a
general relation between the strength of the washout scatterings which
tend to erase any lepton number asymmetry present in the thermal bath,
and the CP asymmetries in the decays of the heavy states.  To our
knowledge, in standard cosmologies only models which invoke a resonant
enhancement of the CP
asymmetries~\cite{Pilaftsis:2003gt,Pilaftsis:2004xx,Pilaftsis:2005rv,GonzalezGarcia:2009qd}
can evade the corresponding bound and bring leptogenesis from heavy
particle decays down to a testable scale~\cite{Chun:2017spz}.

In this paper we point out that this conclusion can be avoided if in
the very early stages, the cosmological history of the Universe is
described by a scalar-tensor gravity (ST)
theory~\cite{jordan1955schwerkraft,Fierz:1956zz,Brans:1961sx} rather
than by general relativity (GR).  ST theories benefit from an
attraction mechanism which, prior to Big Bang Nucleosynthesis (BBN), makes them flow towards
standard GR~\cite{Bartolo:1999sq}, so that discrepancies with direct
cosmological observations can be avoided. 
Such a possibility was already put forth in
relation to possible large enhancements of the dark matter (DM) relic
density with respect to a standard cosmological
evolution~\cite{Catena:2004ba,Catena:2009tm}, as a consequence of a
modified expansion rate.
%
%
Many studies on conformally coupled ST models have been performed
in \cite{Gelmini,RG,WI,LMN,Pallis,Salati,AM,IC,MW2,MW3,Meehan:2015cna,Dutta:2016htz}. In~\cite{Meehan:2015cna} it was 
shown that, in comformally coupled ST theories, BBN constraints can be severe and allow
only moderate deviations from the standard GR expansion history at the
time of DM decoupling so that, based on the boundary conditions and DM masses used in ~\cite{Catena:2004ba,Catena:2009tm},
the enhancement of the DM relic density cannot exceed a factor of three . However, larger DM
masses ($\sim 1$ TeV) were considered in ref.~\cite{Dutta:2016htz} where   
it was shown that the particles undergo a second annihilation process. In that case the
enhancement of the relic density can be   quite large compared to the standard 
cosmological evolution. In ref.~\cite{Dutta:2017fcn}, generalized ST theories were studied after including also disformal
couplings~\cite{Bekenstein:1992pj} and it was concluded that the relic density can be large  even for smaller DM masses, 
with an increase up to a few orders of magnitude compared to the  standard GR case. The larger relic density  in both 
conformal and disformal cases is due to the boosted cosmological expansion rate which characterizes ST theories. 
The magnitude of the enhancement depends  on  boundary conditions  and it can 
be  a few orders of magnitude larger compared to the standard GR expansion. We will make use of this 
enhancement in the expansion rate in our study of leptogenesis.

Differently from the case of DM, for which the typical decoupling
temperature falls in the few GeV range, the generation of a baryon asymmetry via leptogenesis must occur above the EW scale before the EW sphaleron processes get out of equilibrium. 
Since a leptogenesis scale
up to a couple of TeV might still be within the reach of collider
tests, we are interested in modifications of the standard cosmology
at temperatures in the range $100\,{\rm GeV}-{\rm few}\,{\rm TeV}$.  Indeed,
due to the larger scale in the game, we find that in the framework of
conformally coupled ST theories the modified expansion does 
allows to lower the scale for successful leptogenesis down to
$M_1 \lsim 1\,{\rm TeV}$. Hence, in our analysis we will mainly focus on conformally coupled ST theories 
since this conclusion holds also for ST theories with disformal couplings.

The paper is organized as follows. In sec.~\ref{sec:constraint}, we discuss constraints on the leptogenesis scale, in sec.~\ref{sec:expansion}, we discuss cosmological scenarios yielding boosted expansion rates. In sec.~\ref{sec:model} we present a simple benchmark model which will be used for the leptogenesis analysis. In  sec.~\ref{sec:boltzmann}, we discuss the 
network of Boltzmann equations (BE) for leptogenesis in the modified cosmology,  in sec.~\ref{sec:results} 
we discuss our results, in sec.~\ref{sec:mssm}, we discuss the enhancement scales in the extensions of the SM, e.g., Minimal Supersymmetric Standard Model (MSSM) and we conclude in sec.~\ref{sec:conclusions}.    


\section{Constraints on the leptogenesis scale}
\label{sec:constraint}
The quantum field theory conditions required in order that loop
diagrams can generate a lepton ($L$) number (or any other global quantum
number) violating CP asymmetry in the decays of an heavy state $X$
are: (i) complex couplings between $X$ and the particles running in
the loop (say $Y$ and $Z$); (ii) a CP even phase from the loop
factors, which only arises if the $Y,Z$ propagators inside the loops
can go on-shell; (iii) $L$ violation inside the loop.  Condition (ii)
then implies that $Y$ and $Z$ can also participate as external
asymptotic states in scattering processes, and condition (iii) implies
that these scatterings are necessarily $L$ violating. This means that
decay CP asymmetries unavoidably imply $L$ violating washout
scatterings~\cite{Racker:2013lua,Sierra:2013kba}. Since the same
couplings enter both in the expression for the CP asymmetries and for
the washout scattering rates, it is not surprising that a quantitative
relation between CP asymmetries and scattering rates can be worked
out. A general expression for this relation has been obtained
in~\cite{Sierra:2013kba} and reads:
\begin{equation}
  \label{eq:2}
\Gamma(YZ \leftrightarrow  \bar Y\bar Z) \approx \frac{64}{\pi}
\,T\,\left(\frac{T}{M_X}\right)^n \, \epsilon_X^2\,,  
\end{equation}
where $\Gamma$ is the rate of the $\Delta L =2$ washout scatterings,
$\epsilon_X$ is the CP asymmetry in $X$ decays, and $n=0$ for a scalar
$X$ decaying into two scalars $Y,Z$; $n=2$ for a fermion $X$ decaying
into a fermion and scalar pair, and $n=4$ for a scalar $X$ decaying into two fermions. 
Note that \eqn{eq:2} relates scattering washout
rates to CP asymmetries without any reference to the cosmological
model. In relation to successful leptogenesis, cosmology enters
through the requirement that at the relevant temperature $T\sim M_X$
the washout rates do not attain thermal equilibrium:
\begin{equation}
  \label{eq:3}
  \Gamma(YZ \leftrightarrow  \bar Y\bar Z) \lsim \tilde
  H(T)\big|_{T\sim M_X}\,, 
\end{equation}
where we parametrize the deviations of the expansion in terms of a
temperature dependent function $\xi(T)$ multiplying the canonical GR
expansion rate $H(T)= 1.66 \sqrt{g_*}\, T^2/M_P$ (with $g_\star$ the 
relativistic degrees of freedom and $M_P = 1.22 \times 10^{19}$ GeV), namely
$\tilde H(T)= \xi(T)H(T)$ (two examples of $\xi(T)$ are given
in Figure~\ref{fig:speedup})\,.

In the relevant temperature range $T\sim M_X$ the out-of-equilibrium
condition \eqn{eq:3} yields:
\begin{equation}
  \label{eq:generic_constraint}
  M_X \gsim 1.2 M_P \frac{\epsilon^2_X}{\xi(M_X)}\,. 
\end{equation}
Assuming as a benchmark value
$\epsilon_X \gsim 10^{-7}$ as the lowest possible CP asymmetry able to
explain $n_b/n_\gamma\sim 10^{-10}$, we see that standard cosmology
with $\xi(T) = 1$ yields the (conservative) limit
$M_X \gsim 1.4\cdot 10^{5}\,$GeV, so that in any generic model of
leptogenesis from heavy particle decays the relevant scale lies well
above experimental reach.  As an example, we see from the left plot of Figure
\ref{fig:speedup} that in modified ST cosmologies the function
$\xi(T)$ can remain of order $10^2$ in an interval centered at
$T \sim$ TeV and spanning about two orders of magnitude in $z={\rm TeV}/T$ . 
Because of the boosted expansion, in the relevant temperature range, the dynamical 
processes that govern leptogenesis, in particular, the $\Delta L = 2$ washout processes discussed above,
can more easily go out of equilibrium, rendering viable
scales as low as $M_X \lesssim\; $TeV for which direct tests can be foreseen.


\section{Cosmological expansion in scalar-tensor theories}
\label{sec:expansion}
In ST  theories, the gravitational interaction is mediated by both the metric and a scalar field.
The cosmological evolution deviates from the standard expansion of the Universe at early times, but an attractor mechanism ~\cite{Damour:1992kf,Damour:1993id} relaxes the theory towards GR prior to the onset of BBN.
ST theories are often formulated in one of two frames of reference, namely, the Jordan or Einstein frames. As is shown in~\cite{Bekenstein:1992pj}, the most general transformation, physically consistent, between the two metrics of these frames is given by

\begin{equation}\label{metric}
\tilde g_{\mu\nu}=C(\phi)g_{\mu\nu}+D(\phi)\partial_\mu\phi\partial_\nu\phi ,
\end{equation}
where $\tilde g_{\mu\nu}$ is metric in the Jordan frame, $C(\phi)$  is the conformal coupling, $ g_{\mu\nu}$ is the metric in the Einstein frame  and $D(\phi)$ is the so-called disformal coupling. The conformal coupling characterizes the Brans-Dicke class of ST 
theories~\cite{Catena:2004ba,Catena:2009tm,Gelmini,Meehan:2015cna,WI} and the disformal coupling arises naturally in D-brane models, as 
discussed in~\cite{Koivisto:2013fta}.

The respective advantages of these two frames is that the scalar couplings enter through either the gravitational sector (Jordan frame) or the matter sector (Einstein frame), leaving the other sector unaffected. In the Jordan frame matter fields $\Psi$ are coupled directly to the metric, $\tilde g_{\mu\nu}$, which means that the matter sector of the action can be written as $S_{Matter}=S_{Matter}(\tilde g_{\mu\nu},\Psi)$. Thus, this frame is more convenient for particle physics considerations because the usual observables, e.g. a mass, have their standard interpretation. However, the scalar field couples to the gravitational sector producing a rather cumbersome gravitational field equations.

 On the other hand, in the Einstein frame, the matter piece of the action becomes  $S_{Matter}=S_{Matter}(g_{\mu\nu},\phi,\partial_\mu\phi,\Psi)$. This implies that physical quantities associated with particles (i.e. mass) measured in this frame have a spacetime dependency. However, the gravitational field equations take their standard form, where the Einstein tensor is proportional to the total energy momentum tensor.  

The most common strategy followed in the literature ~\cite{Meehan:2015cna,Catena:2004ba,Dutta:2016htz,Dutta:2017fcn} is to determine the cosmic evolution in the Einstein frame, where the cosmological equations take a more straightforward form, and then transform the results over to the Jordan frame. As was already hinted out before, the effect of modified gravity will enter the computation of particle physics processes through the expansion rate, $\tilde H$, in the Jordan frame.  Thus, for our leptogenesis analysis, we implement the standard BE by including a modified Hubble parameter $\tilde H$. In the following paragraphs we will recall some key definitions and we present the equations for ST theories developed in~\cite{Dutta:2016htz,Dutta:2017fcn} that allow to evaluate $\tilde H$.  

The action we consider, written in the Einstein frame, is given by

\begin{eqnarray}\label{action}
 S=\frac{1}{2\kappa^2} \!\!\int{\!d^4x\sqrt{-g}\,R}
- \!\!\int{\!d^4x\sqrt{- g} \left[\frac{1}{2} (\partial\phi)^2+V(\phi)\right]}\nonumber\\
- \!\!\int{\!d^4x\sqrt{-\det\left[ C(\phi)g_{\mu\nu}+D(\phi)\partial_\mu\phi\partial_\nu\phi\right]} \,{\cal L}_{M}(\tilde g_{\mu\nu}) } \,. 
\end{eqnarray}
where $\kappa^2\approx 8\pi G$ (see \cite{Dutta:2016htz}), $\cal{L}_M$  is the matter Lagrangian and $V(\phi)$ is the potential of the scalar field. 

After considering an homogeneous and isotropic FRW metric $g_{\mu\nu}$,
\begin{equation}
ds^2 = - dt^2 + a(t)^2 dx_i dx^i  \,,
\end{equation}
where $a(t)$ is the scale factor, the gravitational field equations and the equation of motion for the scalar field become
\begin{eqnarray}
&& H_E^2 =\frac{\kappa^2}{3} \left[\rho_\phi +\rho\right]\,, \label{friedmann1}\\
&& \dot H_E + H_E^2 = -\frac{\kappa^2}{6}\left[ \rho_\phi+ 3P_\phi +\rho +3 P \right]\,,\label{friedmann2}\\
&& \ddot \phi +3H_E\dot\phi +  \frac{d V}{d \phi}+Q_0 =0 \label{eomfield}\,,
\end{eqnarray}
In the previous equations, dots represent derivatives with respect to the time in the Einstein frame, $H_{E}\equiv\frac{\dot a}{a}$ is the expansion rate in the Einstein frame. Note that $H_{E}$ is not the same as the expansion rate in standard cosmology, which we denote by $H$. Additionally, $\rho$ and $P$ are the energy density and pressure of the universe written in the Einstein frame. Moreover, the energy density and pressure of the scalar field are $\rho_\phi=\frac{1}{2}\dot\phi^2+V(\phi)$ and $P_\phi=\frac{1}{2}\dot\phi^2-V(\phi)$. 

In \eqref{eomfield} we introduce $Q_0$, which is given by

\begin{equation*}
Q_0 = \rho \left[ \frac{D}{C} \,\ddot \phi + \frac{D}{C} \,\dot \phi \left(\!3H_E + \frac{\dot \rho}{\rho} \right) \!+ \!\left(\!\frac{D_{,\phi}}{2C}-\frac{D}{C}\frac{C_{,\phi}}{C}\!\right) \dot\phi^2 +\frac{C_{,\phi}}{2\,C} (1-3\,\omega)
\right], \\
\end{equation*}
where $\omega=P/\rho$ is measured in the Einstein frame. $\omega$ can be related to the equation of state parameter in the Jordan frame ($\tilde \omega$) , which is the frame where temperature takes the standard interpretation, through $\omega = (1+D C^{-1}\dot\phi^2)  \tilde\omega$. To calculate $\tilde \omega$ in the very early universe, one has to consider the contribution of each particle in the cosmic fluid to the energy density and pressure of the universe. Throughout most of the early radiation era $\tilde \omega=1/3$, but once the temperature of the universe drops below the rest mass of each particle, $\tilde \omega$ becomes slightly less than one third.  

In order to solve the cosmological equations, it is convenient to replace time derivatives of a generic function $f$ 
with derivatives with respect to the number of e-folds N ($dN=Hdt$), which will be denoted with a prime $f'=df/dN$. We also introduce a dimensionless scalar field $\varphi=\kappa\phi$ for convenience.

After combining \eqref{friedmann1}, \eqref{friedmann2} and \eqref{eomfield} one arrives to the so-called master equation, which describes the evolution of the scalar field during any epoch of the universe (see ref.~\cite{Dutta:2016htz}). During the  radiation dominated era, $\rho$ dominates over $V$, so that we can take $V\approx 0$. Moreover, in this work, we will focus on the pure conformal case and set $D=0$.  Under those considerations, the master equation becomes 
\begin{equation}\label{mastereq}
 \frac{2}{1-\varphi'^2/6} \,\varphi'' + \left(1 -\omega \right)\varphi'  +2 (1 -3\,\omega)\,\alpha(\varphi)=0\,, 
\end{equation}
where $\alpha(\varphi)=\frac{d\ln C^{1/2}}{d\varphi}$. We consider the conformal coupling $C(\varphi)=1+0.1\exp[-8\varphi]$, which has been used in previous works \cite{Catena:2004ba,Dutta:2016htz,Dutta:2017fcn}.  

As was stated earlier, we need the expansion rate in the Jordan frame for our leptogenesis calculation. This expansion rate, $\tilde H$, can be written as (see ~\cite{Dutta:2016htz})
\begin{equation}\label{Htilde}
\tilde H^2 =  \frac{\kappa^2}{\kappa_{GR}^2}\frac{C (1+\alpha(\varphi) \varphi')^2}{1-\varphi'^2/6}\, H^2 \,,
\end{equation}
where $H^2=\frac{\kappa^2_{GR}}{3}\tilde\rho$, $\kappa^2\simeq\kappa^2_{GR}=8\pi G$ and $\tilde \rho\sim g(T) T^4$ for the radiation dominated era. From this relation the speedup parameter $\xi$ can be defined as
\begin{equation}
\xi\equiv\frac{\tilde H}{H} \,.
\end{equation} 

As mentioned above, the evolution of the scalar field is described by \eqref{mastereq}. This equation contains a term that can be interpreted as an effective potential, given by $V_{eff}=\ln C$. During the radiation dominated era, $\omega=1/3$ and the effective potential term disappears. Later on, around the time when the particles of the plasma become non-relativistic, the parameter $\omega$ in the 
equation of state differs slightly from $1/3$ (see \cite{Dutta:2016htz}) and the effective potential kicks in.  Therefore, the evolution of the scalar field depends on the effective potential, the initial conditions chosen, and the particle content. 

In general, both the initial position and velocity of the scalar field can take any positive or negative values. For the conformal factor chosen, we see that if the velocity is positive or zero, the scalar field will roll down the potential and will slow down due to Hubble friction to a final positive value. That is, the conformal factor will evolve rapidly towards 1, and hence the modification to the expansion rate would be negligible (see eq.~\eqref{Htilde}). 

\begin{figure}[h!]
\centerline{
\begin{tabular}{cc}
\includegraphics[width=0.48\textwidth]{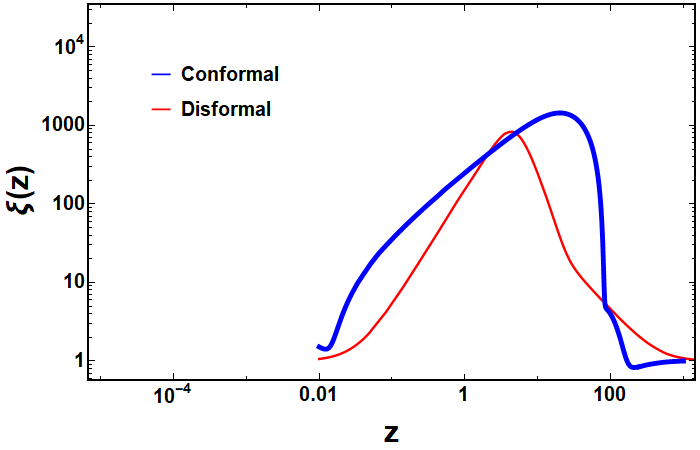}&\includegraphics[width=0.48\textwidth]{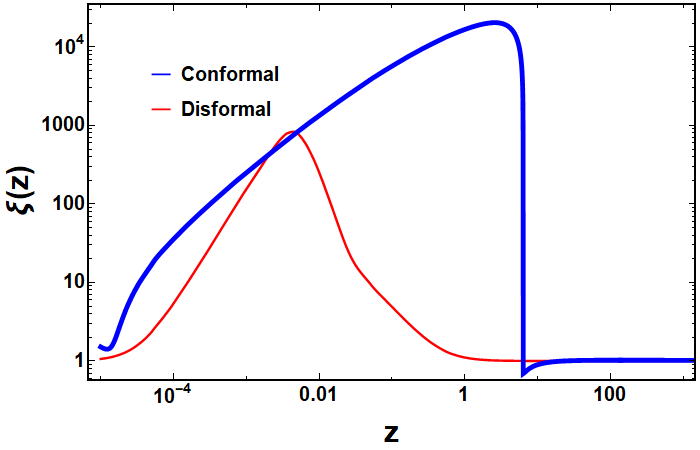}
\end{tabular}}
\caption{Two examples for the speedup factor $\xi(z)$ for conformal (thick blue lines) and disformal (thin red lines)  scenarios, as a function of $z=1\,\rm{TeV}/T$ with $T$ the temperature. Left plot: $\xi(z)$ in a conformal scenario with initial conditions $(\varphi_i,\varphi'_i)=(0.8,-1.83)$ and initial temperature  $100\,$TeV, and in a disformal scenario with $(\varphi_i,\varphi'_i)=(0.2,-2\times 10^{-6})$ and a mass scale $M_D=2.5\,$TeV. Right plot: $\xi(z)$ in a conformal and disformal scenario with  initial conditions as before, but initial temperature $T=10^5\,$TeV, and $M_D=2500\,$TeV for the disformal case.}
\label{fig:speedup}
\end{figure}

A more interesting result arises when considering negative velocities of the scalar field. In this case, the field will start rolling-up the effective potential towards smaller values of the field, eventually turning back down and moving towards its final value. So, if the field starts at a positive value, and given a sufficiently negative initial velocity, it will move towards negative values until its velocity becomes zero and then positive again, as it rolls back down the effective potential. This change in sign for the scalar field will produce a peak in the conformal coupling, which will give rise to a non-trivial modification of the expansion rate $\xi\neq 1$. This particular behavior is shown in Figure \ref{fig:speedup}.

The equation of state parameter $\omega$ plays an important role in locating the temperature at which the speedup factor drops back to 1. 
Slight variations from the radiation dominated value $\omega=1/3$ appear when particles become non-relativistic. So, to calculate $\omega$, one has to take into account all the SM particles and, depending on the specific SM extensions one is dealing with,  one would add right-handed neutrinos, supersymmetric partners or other type of heavy species.  

In choosing the boundary conditions to solve eq.~\eqref{mastereq}, care must be taken to respect   
the constraints imposed by the post-Newtonian parameters~\cite{Bertotti:2003rm,Shapiro:2004zz,Williams:1995nq,Will:2001mx,EspositoFarese:2000ij} and, 
most importantly,  one has to ensure that by the time of the onset of BBN $\xi\approx1$. 

Another interesting scenario yielding modified Hubble parameters is the pure disformal scenario, which is defined by a conformal coupling equal to one, and a disformal coupling different from zero in eq.~\eqref{metric}. This scenario is studied extensively in \cite{Dutta:2017fcn} where the authors present the mathematical formalism, solve the necessary equations for the evolution of the scalar field, and find the modifications to the expansion rate for the particular case $D=1/M_D^4$, where $M_D$ is a mass scale motivated by String Theory, and depends on the string coupling and string scale. 

In Figure \ref{fig:speedup} we also present the speedup factor $\xi$ in a pure disformal case (thin red lines). In this scenario, $M_D$ plays the most important role in the location and shape of $\xi$. The maximum $\xi$ happens close to a temperature equal to $M_D$. It is interesting to notice that by rescaling $M_D$, $\xi$ moves to a higher (or lower) temperature without changing shape.


Due to the fact that, as in GR, also in ST cosmologies the total entropy is conserved, adapting the BE for leptogenesis 
to ST cosmologies is rather straightforward. Considering just the BE for the evolution of the RH neutrinos 
density $n_N$ including only decays and inverse decays will suffice to illustrate this. Denoting 
with $a$ the scale factor, the BE reads:
\begin{equation}
  \label{eq:BE1}
  a^{-3}\frac{d(n_N a^3)}{d t} = \langle \Gamma_N\rangle (n_N^0-n_N)\,,
\end{equation}
where  $\langle \Gamma_N\rangle$ is the thermal averaged decay rate and $n_N^{\rm eq}$ the 
equilibrium density. As usual we use the fiducial variable $z=M_N/T$ with $T$ the temperature, and write the time derivative as:
\begin{equation}
\label{eq:BE2} 
\frac{d}{d t} = z \left(\frac{1}{z}\frac{d z}{dt}\right) \frac{d}{dz}
= z  \left(\frac{1}{a}\frac{d a}{dt}\right)\frac{d}{dz} = z \tilde H \frac{d}{dz}\,.
\end{equation}
The second step relies on entropy conservation $d(s a^3)/dt = 0$
with $s\propto T^3$ the entropy density which implies the usual 
temperature-scale factor relation  $T\propto 1/a$,  
while $\tilde H$ is the physical Hubble parameter
defined as the rate of change of the physical length scale.
The rest is standard: denoting by $\gamma _N = n_N^{\rm eq}  \langle \Gamma_N\rangle$ 
the density of the reaction, and normalizing the particle number
densities to the entropy density as $Y_N = n_N/s$ we have:
\begin{equation}
  \label{eq:BE3}
  \frac{d Y_N }{d z} = 
\frac{1}{  s z \tilde H}  \left(1- \frac
  {Y_N}{Y_N^{\rm eq}}\right)\gamma_N=
\frac{1}{  s z H}  \left(1- \frac    {Y_N}{Y_N^{\rm eq}}\right)\frac{\gamma_N}{\xi(z)}
\,.
\end{equation}
In the second equation we have rewritten $\tilde H = H \, \xi(z)$
with $H= H_{GR}$  and 
$\xi (z)>1$ the $T$-dependent speedup factor, to put in evidence
how in the BE its effect is equivalent to ``slowing down'' the decay
and inverse decay reactions, favoring the enforcement of the
out-of-equilibrium condition.

In the next section, we will show how a simple (non-resonant) leptogenesis model, when embedded in a non-standard 
cosmology characterized by  a boosted expansion rate,   allows to get around the constraint \eqn{eq:generic_constraint}. It is interesting to remark that if a particle physics model can be experimentally established as responsible for the 
cosmological baryon asymmetry via (non-resonant)  baryogenesis via decays of TeV scale particles, this would constitute a direct 
evidence of non-standard cosmology in a temperature range unreachable by all other cosmological probes 
(DM freeze-out, EW phase transition, etc.).

\section{A simple test model}  
\label{sec:model}

Besides the generic constraint in \eqn{eq:generic_constraint}, the type-I (non-resonant) leptogenesis is subject to \eqn{eq:DIbound} from neutrino mass. In order to get around the latter, a simple way is to assume that the VEV responsible for the neutrino masses is much smaller than the full EW breaking VEV: $v_\nu \ll v\sim 174\,$GeV.  
By requiring a sufficient CP
asymmetry $\epsilon_1 \gtrsim 10^{-7}$, the scale $M_1 \sim 1\,$TeV can be
reached for $v_\nu \lsim 0.2\,$GeV.  Of course one has to introduce an
{\it ad hoc} Higgs field with $\langle H_\nu \rangle = v_\nu$ coupled
to RH neutrinos $N_j$ as $\lambda_{j\alpha } \bar N_j L_\alpha H_\nu$,
and forbid the couplings with the standard Higgs $H$ via some $\zb_2$
or $U(1)$ symmetry (a $U(1)$ softly broken might be preferrable to
avoid domain wall problems with a spontaneously broken $\zb_2$).  Such
model exists, see for example~\cite{Haba:2011ra},
or~\cite{Clarke:2015hta} for various different possibilities (in the
last paper, Model Type I with $m^2_{12}>0$ and $\lambda_5=0$ is
probably the best option).  Although the `neutrinophilic' VEV model (we will denote it as $v_\nu$-model)
might not represent the most elegant possibility, its structure
remains very similar to the standard type-I see-saw model, with the
advantage that it minimizes the differences with respect to the
standard leptogenesis case, rendering it suitable as test model to
illustrate the effects of non-standard cosmologies.\footnote{Some
  $\Delta L=1$ $2 \leftrightarrow 2$ washouts involving the top quark,
  like $Q_{3L} L \leftrightarrow N t_R$ and
  $Q_{3L} \bar t_R \leftrightarrow N \bar L$ will be absent, since
  $H_\nu$ does not couple to the top-quark. This has no major impact
  in determining the viable leptogenesis scale.}
The usual seesaw formula still holds:
\begin{equation}
  \label{eq:seesaw}
m_\nu \simeq \lambda^T\frac{v^2_\nu}{M}\lambda,
\end{equation}
and so does the Casas
Ibarra parametrization of the Yukawa couplings:
\begin{equation}
  \label{eq:CI}
  \lambda_{j\alpha} = \frac{1}{v_\nu} \sqrt{M_j}R_{j\beta}
  \sqrt{m_\beta} (U^\dagger)_{\beta\alpha}\,, 
\end{equation}
with $M_j$ and $m_\beta$ the heavy and light neutrino mass eigenvalues, $U$ the
neutrino mixing matrix, and $R$ a generic complex orthogonal matrix ($RR^T=I$).

Let us consider \eqn{eq:seesaw}. In the usual seesaw with the SM VEV
$v\sim 174\,$GeV, to allow for a low value of $M$ while still ensuring
$m_\nu \lsim 0.1\,$eV, one has to take tiny Yukawa couplings $\lambda$, which in turn imply tiny CP asymmetries.\footnote{One could
  arrange for cancellations in the matrix multiplications to
  keep the coupling $\lambda$
  sizable~\cite{Hambye:2003rt,Raidal:2004vt}. However, 
  this requires exponential fine tunings in the phases of the complex angles of the $R$ matrix in 
  \eqn{eq:CI}~\cite{Ibarra:2010xw,Shuve:2014zua} which, moreover, 
  are unstable under quantum corrections~\cite{AristizabalSierra:2011mn}.}
In the $v_\nu$-model instead, the couplings $\lambda$ can be large
since it is $v_\nu$ that is small, and thus the CP asymmetries can be
also large.  However, if the scale $M_1$ is low, leptogenesis will
occur when the Universe expansion is slower, and then the
$\Delta L =2$ washouts $L H \leftrightarrow \bar L \bar H$ or
$LL\leftrightarrow \bar H \bar H$ that are mediated by the same Yukawa
couplings can attain thermal equilibrium, realizing the situation in which
leptogenesis cannot be successful because of the constraint discussed
in sec.~\ref{sec:constraint}.

\begin{figure}[t!!]
\begin{center}
\includegraphics[width=0.8 \textwidth]{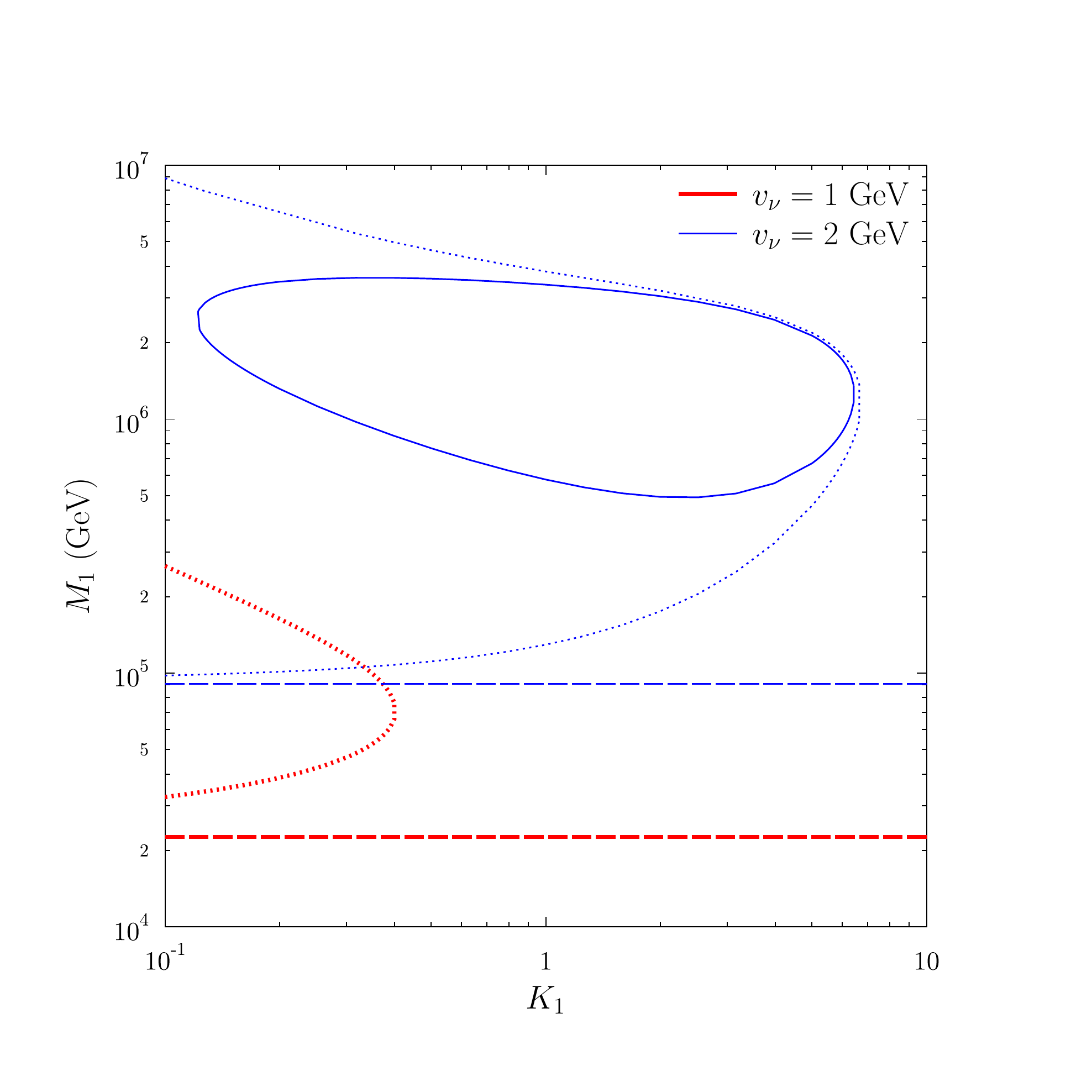}
\end{center}
\caption{The bounds on $M_1$ for $v_\nu = 1, 2$ GeV (red/thick, blue/thin lines) as a function of $K_1$ defined in \eqn{eq:K_1}. Outside these closed regimes, one cannot obtain sufficient baryon asymmetry. The solid lines are for zero initial abundance of $N_1$ while dotted lines for thermal initial abundance of $N_1$. The horizontal dashed lines are the absolute lower bounds obtained for the respective $v_\nu$ which correspond to having thermal initial abundance of $N_1$ and no washout.}
\label{fig:GR_bound_M1}
\end{figure}

In order to illustrate these constraints in the GR, we show in Figure \ref{fig:GR_bound_M1} the bounds on the lightest RH neutrino mass $M_1$ from leptogenesis in the $v_\nu$-model as a function of \emph{washout} parameter defined as 
\be
K_1 = \left.\frac{\Gamma_{N1}}{H}\right|_{T = M_1}
\label{eq:K_1}
\ee
where $\Gamma_{N_1} = \frac{(\lambda\lambda^\dagger)_{11} M_1}{8\pi}$ is the total decay width of $N_1$. Outside these regions, one cannot generate sufficient baryon asymmetry.
Notice that for $K_1 = 1$, the \emph{effective} neutrino mass is
\be
{m_{\rm eff}} \equiv 
\frac{(\lambda\lambda^\dagger)_{11} v_\nu^2}{M_1} 
= 3.6\times 10^{-8} 
\sqrt{\frac{g_\star}{110.75}}
\left(\frac{v_\nu}{1\,{\rm GeV}}\right)^2\,{\rm eV}.
\label{eq:m_eff}
\ee
This implies that even in the strong washout regime $K_1 \gg 1$, 
the lightest light neutrino remains essentially massless i.e. $m_1 \approx 0$.

In Figure \ref{fig:GR_bound_M1}, the red/thick and blue/thin lines correspond respectively to values of $v_\nu = 1, 2$ GeV.\footnote{The results are obtained from solving eqs. \eqref{eq:BE_N1} and \eqref{eq:BE_Delta_alpha_sim} by setting $\xi = 1$ and the heaviest neutrino mass $m_3 = 0.05$ eV. For further details, refer to sec.~\ref{sec:boltzmann}.} 
The solid lines are for the case of vanishing initial $N_1$ abundance $Y_{N_1}(0)=0$ while dotted lines for thermal initial $N_1$ abundance $Y_{N_1}(0)=Y_{N_1}^{\rm eq}(0)$. 
The horizontal dashed lines refer to the absolute lower bounds obtained for the respective $v_\nu$ if leptogenesis proceeds with $Y_{N_1}(0)=Y_{N_1}^{\rm eq}(0)$ in the absence of washout (or in the weak washout regime $K_1 \ll 1$). 
The lower and upper bounds are respectively due to \eqn{eq:DIbound} and $\Delta L = 2$ washout scattering discussed in sec. \ref{sec:constraint}. Notice that one cannot lower the scale of $M_1$ indefinitely by lowing $v_\nu$, at some point, the washout will be too strong to generate sufficient baryon asymmetry. This is the case for $v_\nu = 1$ GeV where no solution exists for the case of $Y_{N_1}(0)=0$. In this case, one arrives at lower bound on $M_1$ of few times $10^5$ GeV in agreement with the estimation in sec. \ref{sec:constraint}.\footnote{Our results for $Y_{N_1}(0)=Y_{N_1}^{\rm eq}(0)$ are also consistent with the estimation in refs.~\cite{Haba:2011ra,Clarke:2015hta}.}   

\section{Boltzmann equations in the modified cosmology}
\label{sec:boltzmann}

In the following, we will describe a particle $X$ in term of abundance $Y_X \equiv n_X/s$ defined as its number density $n_X$ normalized over entropic density $s = \frac{2\pi^ 2}{45}g_\star T^3$. In the following, we will fix $g_\star = 110.75$ for the SM with an additional (neutrinophilic) Higgs doublet. We start with the following BEs for $Y_{N_{1}}$ and
$Y_{\Delta_{\alpha}}$ with $\Delta_{\alpha}\equiv\frac{B}{3}-L_{\alpha}$
as follows\footnote{
To avoid double counting in the BE
for $Y_{\Delta_{\alpha}}$, we have subtracted off the CP-violating
$\Delta L=2$ scattering involving on-shell $N_{1}$ and ignored the
off-shell contribution \cite{Kolb:1979qa,Fry:1980ph}.}
\begin{eqnarray}
s\xi Hz\frac{dY_{N_{1}}}{dz} & = & -\gamma_{N_{1}}\left(\frac{Y_{N_{1}}}{Y_{N_{1}}^{{\rm eq}}}-1\right), 
\label{eq:BE_N1} \\
s\xi Hz\frac{dY_{\Delta_{\alpha}}}{dz} & = & -\epsilon_{1\alpha}\gamma_{N_{1}}\left(\frac{Y_{N_{1}}}{Y_{N_{1}}^{{\rm eq}}}-1\right)+\frac{1}{2}P_{1\alpha}\gamma_{N_{1}}\left(\frac{Y_{\Delta\ell_{\alpha}}}{Y_{f}}+\frac{Y_{\Delta H_{\nu}}}{Y_{b}}\right)\nonumber \\
 &  & +\gamma_{22}^{\alpha\alpha}\left(\frac{Y_{\Delta\ell_{\alpha}}}{Y_{f}}+\frac{Y_{\Delta H_{\nu}}}{Y_{b}}\right)+\sum_{\beta\neq\alpha}\gamma_{22}^{\alpha\beta}\left(\frac{Y_{\Delta\ell_{\alpha}}}{2Y_{f}}+\frac{Y_{\Delta\ell_{\beta}}}{2Y_{f}}+\frac{Y_{\Delta H_{\nu}}}{Y_{b}}\right),
\label{eq:BE_Delta_alpha}
\end{eqnarray}
where we have defined $z\equiv\frac{M_{1}}{T}$, $Y_{f}\equiv\frac{15}{8\pi^{2}g_{\star}}$
and $Y_{b}\equiv\frac{15}{4\pi^{2}g_{\star}}$. 
In the above, $Y_{\Delta\ell_{\alpha}}$
and $Y_{\Delta H_{\nu}}$ refer to abundances per gauge degrees of
freedom. Explicitly, the total thermal averaged decay reaction density $\gamma_{N_{1}}$ is given by
\be
\gamma_{N_1} =
\sum_{\alpha}\gamma_{N_{1}\to\ell_{\alpha}H_{\nu}} 
= n_{N_1}^{\rm eq} \Gamma_{N_1} \frac{{\cal K}_1(z)}{{\cal K}_2(z)},
\label{eq:gamma_N_1}
\ee
where ${\cal K}_n(z)$ refers to the modified Bessel function of second type of order $n$ and the branching ratio for $N_1$ decay to lepton of flavor $\alpha$ as $P_{1\alpha}\equiv\frac{\gamma_{N_{1}\to\ell_{\alpha}H_{\nu}}}{\gamma_{N_{1}}}$.
The $\Delta L=2$ washout mediated by off-shell $N_{i}$ is described by $\gamma_{22}^{\alpha\beta}$.\footnote{
The $\Delta L=1$ scatterings involving gauge bosons are not considered since to consider them consistently, one also needs to consider CP violation in them which will result in a small net
effect \cite{Nardi:2007jp,Fong:2010bh}. As for flavor changing but $\Delta L=0$ scatterings, their rates go as $\frac{T^{5}}{M_{i}^{4}}$
for $T<M_{i}$ which are less important than that of $\Delta L=2$ reactions which go as $\frac{T^{3}}{M_{i}^{2}}$ for $T<M_{i}$. Furthermore, in the following, we will consider democratic flavor structure where
they are either not relevant or in thermal equilibrium and can be
dropped from the BEs. } In order to minimize the complication from flavor effects and focus solely on the effect of ST cosmology, we choose the democratic flavor structure as follows
\begin{eqnarray}
\epsilon_{1e} & = & \epsilon_{1\mu}=\epsilon_{1\tau}\equiv\frac{\epsilon_{1}}{3},\\
P_{1e} & = & P_{1\mu}=P_{1\tau}=\frac{1}{3},\\
\gamma_{22}^{\alpha\beta} & \equiv & \frac{1}{9}\gamma_{22},
\end{eqnarray}
where $\epsilon_{1}\equiv\sum_{\alpha}\epsilon_{1\alpha}$ and $\gamma_{22}\equiv\sum_{\alpha\beta}\gamma_{22}^{\alpha\beta}$.
With the above assumptions, the BE for $Y_{\Delta_{\alpha}}$
becomes
\begin{eqnarray}
s\xi Hz\frac{dY_{\Delta_{\alpha}}}{dz} & = & -\frac{1}{3}\gamma_{N_{1}}\epsilon_{1}\left(\frac{Y_{N_{1}}}{Y_{N_{1}}^{{\rm eq}}}-1\right)+\frac{1}{6}\gamma_{N_{1}}\left(\frac{Y_{\Delta\ell_{\alpha}}}{Y_{f}}+\frac{Y_{\Delta H_{\nu}}}{Y_{b}}\right)\nonumber \\
 &  & +\frac{1}{9}\gamma_{22}\left(\frac{Y_{\Delta\ell_{\alpha}}}{Y_{f}}+\frac{Y_{\Delta H_{\nu}}}{Y_{b}}\right)+\frac{1}{9}\gamma_{22}\left(\frac{Y_{\Delta\ell_{\alpha}}}{Y_{f}}+2\frac{Y_{\Delta H_{\nu}}}{Y_{b}}+\frac{1}{2}\sum_{\beta\neq\alpha}\frac{Y_{\Delta\ell_{\beta}}}{Y_{f}}\right).\label{eq:BE_Delta_alpha_sim}
\end{eqnarray}

For $M_{2,3}\gg M_{1}$, the total CP parameter is given by \cite{Roulet:1998}
\begin{eqnarray}
\epsilon_{1} & \simeq & -\frac{3}{16\pi}\sum_{j>1}\frac{{\rm Im}\left[\left(\lambda\lambda^{\dagger}\right)_{1j}^{2}\right]}{\left(\lambda\lambda^{\dagger}\right)_{11}}\frac{M_{1}}{M_{j}},
\end{eqnarray}
and using \eqn{eq:CI}, one can derive the Davidson-Ibarra bound \cite{Davidson:2002qv}
\begin{eqnarray}
\left|\epsilon_{1}\right| & \leq & \frac{3}{16\pi}\frac{M_{1}}{v_{\nu}^{2}}\left(m_{3}-m_{1}\right)\equiv\epsilon_{1}^{{\rm max}},\label{eq:DIbound_vnu}
\end{eqnarray} 
as introduced in \eqn{eq:DIbound} but with $v \to v_\nu$.
We further parametrize the off-shell $\Delta L=2$ washout mediated by $N_{i}$ valid for $T<M_{i}$ as follows
\begin{eqnarray}
\gamma_{22} & \equiv & \frac{n}{\pi^{3}}\frac{{\rm Tr}\left[m_{\nu}m_{\nu}^{\dagger}\right]}{v_{\nu}^{4}}T^{3}
=\frac{nM_{1}^{3}}{\pi^{3}z^{3}}\frac{\sum_{i}m_{i}^{2}}{v_{\nu}^{4}},
\label{eq:eff_22_washout}
\end{eqnarray}
where $n=\frac{2}{\pi^{2}}T^3$. As shown in \eqn{eq:m_eff}, the lightest light neutrino mass $m_1$ can be neglected and we can rewrite the above in term of \eqn{eq:DIbound_vnu} as follows
\begin{eqnarray}
\gamma_{22} & \simeq & \frac{256 n}{9\pi} \frac{M_1}{z^{3}}\epsilon_{1}^{{\rm max},2}.
\label{eq:eff_22_washout2}
\end{eqnarray}
%
From the above, we see that the $\Delta L=2$ washout is indeed proportional to $\epsilon_{1}^{{\rm max},2}$ as argued in \eqn{eq:2}, so that $M_1$ remains bounded from below by the general lower limit given in  \eqn{eq:generic_constraint}. 
As in the standard type-I seesaw, also in the present case an upper bound on $M_1$ exists, which follows from the requirement 
that $\Delta L=2$ washout scatterings will not become too strong to erase the asymmetry.
\Eqn{eq:eff_22_washout} shows that once the neutrino mass scale $m_i$ is fixed, for each value of  $v_\nu$ there is a limiting upper value of $M_1$ for which $\gamma_{22}$ remains sufficiently out of equilibrium. However, 
while in the standard case this hints to a loose upper limit of order $\sim 10^{14}\,$GeV, due to the large 
hierarchy $v_\nu/v\lsim 10^{-2}$ and to the quartic dependence on the VEV values, in the neutrinophilic VEV model 
the corresponding constraint is much stronger.

For the spectator effects \cite{Buchmuller:2001sr,Nardi:2005hs}, we consider the temperature regime $T\lesssim 10^{5}$
GeV where all Yukawa interactions are in chemical equilibrium. We further assume that $H_{\nu}$ does not carry a conserved charge\footnote{This can be due to fast interactions induced by $\lambda_{5}\left(H^{\dagger}H_{\nu}\right)^{2}$
in the scalar potential.} and we have
\begin{eqnarray}
\left(\begin{array}{c}
Y_{\Delta\ell_{e}}\\
Y_{\Delta\ell_{\mu}}\\
Y_{\Delta\ell_{\tau}}
\end{array}\right) & = & \frac{1}{207}\left(\begin{array}{ccc}
-64 & 5 & 5\\
5 & -64 & 5\\
5 & 5 & -64
\end{array}\right)\left(\begin{array}{c}
Y_{\Delta_{e}}\\
Y_{\Delta_{\mu}}\\
Y_{\Delta_{\tau}}
\end{array}\right),\\
Y_{\Delta H_{\nu}} & = & -\frac{2}{23}\left(Y_{\Delta_{e}}+Y_{\Delta_{\mu}}+Y_{\Delta_{\tau}}\right).
\end{eqnarray}

Substituting the result above into eq. (\ref{eq:BE_Delta_alpha_sim})
and summing over $\alpha$ on both sides, the BE becomes
\begin{eqnarray}
s\xi Hz\frac{dY_{\Delta}}{dz} & = & -\gamma_{N_{1}}\epsilon_{1}\left(\frac{Y_{N_{1}}}{Y_{N_{1}}^{{\rm eq}}}-1\right)+\frac{1}{6}\gamma_{N_{1}}\left(-\frac{6}{23}\frac{Y_{\Delta}}{Y_{f}}-\frac{6}{23}\frac{Y_{\Delta}}{Y_{b}}\right)\nonumber \\
 &  & +\frac{1}{9}\gamma_{22}\left(-\frac{6}{23}\frac{Y_{\Delta}}{Y_{f}}-\frac{6}{23}\frac{Y_{\Delta}}{Y_{b}}\right)+\frac{2}{9}\gamma_{22}\left(-\frac{6}{23}\frac{Y_{\Delta}}{Y_{f}}-\frac{6}{23}\frac{Y_{\Delta}}{Y_{b}}\right)\nonumber \\
 & = & -\gamma_{N_{1}}\epsilon_{1}\left(\frac{Y_{N_{1}}}{Y_{N_{1}}^{{\rm eq}}}-1\right)-\frac{3}{23}\left(\gamma_{N_{1}}+2\gamma_{22}\right)\frac{Y_{\Delta}}{Y_{f}},
\label{eq:BE_Delta_alpha_sim2}
\end{eqnarray}
where we have defined $Y_{\Delta}\equiv Y_{\Delta_{e}}+Y_{\Delta_{\mu}}+Y_{\Delta_{\tau}}$.
In ref. \cite{DOnofrio:2014rug}, assuming the SM, it was obtained that the EW sphaleron processes freeze out at $T_{\rm EWSp}=132$ GeV after the EW symmetry breaking at $T_c=159$ GeV. Assuming the EW symmetry breaking also happens before $T_{{\rm EWSp}}$ in $v_\nu$-model, we have \cite{Harvey:1990qw,Inui:1993wv}
\begin{eqnarray}
Y_{\Delta B} & = & \frac{30}{97}Y_{\Delta},
\label{eq:Delta_to_B}
\end{eqnarray}
excluding the contributions from heavy charged (neutrinophilic) Higgs and top quark.


\section{Results}
\label{sec:results}

The asymmetry $Y_\Delta$ can be parametrized in term of \emph{efficiency} factor $\eta = \eta(K_1,v_\nu,m_3,M_1)$ as follows
\begin{eqnarray}
Y_{\Delta} & = & \epsilon_{1}Y_{N_{1}}^{{\rm eq},0}\eta,
\end{eqnarray}
where $Y_{N_{1}}^{{\rm eq},0} \equiv Y_{N_{1}}^{{\rm eq}}\left(0\right)=\frac{45}{\pi^4 g_\star}$. 
The above parametrization is convenient because once we substitute it into \eqn{eq:BE_Delta_alpha_sim2}, for temperature-independent $\epsilon_1$, the BE becomes independent of $\epsilon_1$. The final asymmetry is obtained by evaluating the final efficiency $\eta = \eta(z\to\infty)$. In the case with an initial thermal abundance of $N_1$, one saturates to the maximal efficiency $\eta = 1$ in limit of weak washout $K_1 \ll 1$ and small $\Delta L = 2$ washout. As we will see in more detail later, as $M_1$ gets close to the EW sphaleron freezeout temperature $T_{\rm EWSp}$, one might not be able to saturate the efficiency factor because the baryon asymmetry will be frozen before all $N_1$ can decay.

Using \eqn{eq:DIbound_vnu} and \eqn{eq:Delta_to_B}, the maximal asymmetry is given by
\begin{eqnarray}
Y_{\Delta B}^{{\rm max}} & = & \frac{30}{97}\frac{3}{16\pi}\frac{M_{1}m_{3}}{v_{\nu}^{2}}Y_{N_{1}}^{{\rm eq},0}\eta.
\end{eqnarray}
Setting $Y_{\Delta B}^{{\rm max}} = Y_{\Delta B}^{{\rm obs}} = 8.7 \times 10^{-11}$, we can derive both upper and lower bounds on $M_{1}$. Starting from a very small $M_{1}$ while keeping the $\Delta L = 2$ washout under control, the CP parameter might be too small and we need to increase $M_{1}$ until $Y_{\Delta B}^{{\rm max}} = Y_{\Delta B}^{{\rm obs}}$, which gives us the lower bound on $M_{1}$. As we continue to increase $M_{1}$, eventually the $\Delta L=2$ washout \eqn{eq:eff_22_washout} will become too strong until which we are no longer able to obtain sufficient baryon asymmetry and this gives us an upper bound on $M_1$. As we explain below \eqn{eq:eff_22_washout2}, this upper bound is specific to the model we have chosen due to neutrino mass constraint. In other words, from the following equation
\begin{eqnarray}
M_{1}\eta\left(K_{1},v_{\nu},m_3,M_{1}\right) & = & \frac{97}{30}\frac{16\pi}{3}\frac{v_{\nu}^{2}}{m_{3}}\frac{Y_{\Delta B}^{{\rm obs}}}{Y_{N_{1}}^{{\rm eq},0}},
\end{eqnarray}
for a given $v_{\nu}$ and $m_{3}$, we can have no solution, one solution,
or two solutions for $M_{1}$. The two solutions will correspond to 
upper and lower bounds on $M_{1}$. Notice that as we go to smaller $M_1$, the EW sphaleron freezeout temperature becomes relevant and we fix this to be $T_{\rm EWSp} =132$ GeV after which the value of baryon asymmetry will be frozen.

\begin{figure}[t!!]
\begin{center}
\includegraphics[width=0.8 \textwidth]{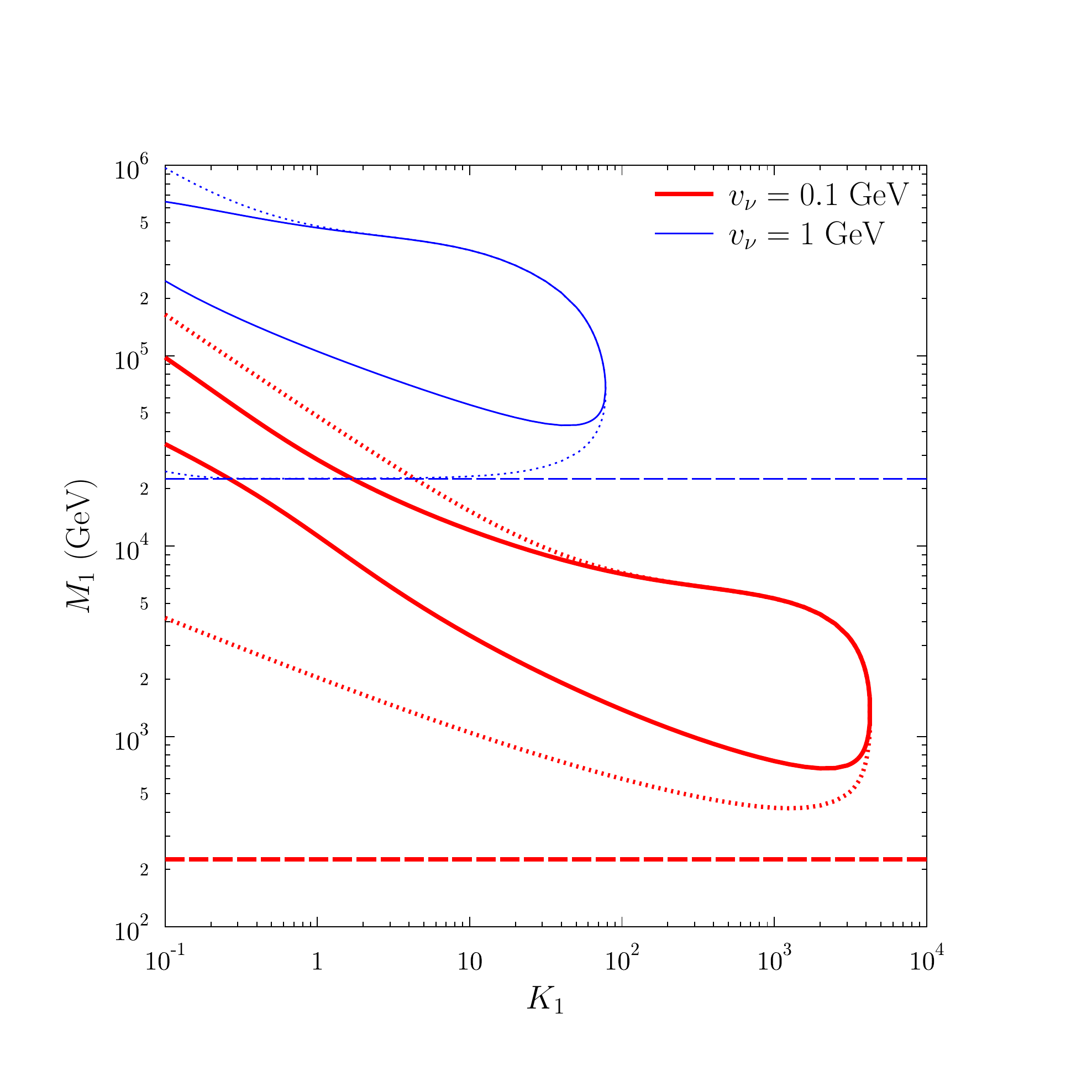}
\end{center}
\caption{The bounds on $M_1$ for $v_\nu = 0.1, 1$ GeV (red/thick, blue/thin lines) as a function of $K_1$ defined in \eqn{eq:K_1}. Outside the closed regimes, one cannot obtain sufficient baryon asymmetry.
The solid lines are for zero initial abundance of $N_1$ while dotted lines for thermal initial abundance. 
The horizontal dashed lines represent the absolute lower bounds on $M_1$ for the respective $v_\nu$ 
obtained with initial thermal abundance of $N_1$ and no washout.}
\label{fig:ST_bound_M1}
\end{figure}

It is important to note the temperature where speedup happens is crucial for leptogenesis. 
As discussed in sec. \ref{sec:expansion}, while the regime of speedup for conformal case depends on initial temperature and $\varphi$ field configurations, the regime of speedup for disformal case depends on a new mass scale $M_D$. 
Besides this point, the qualitative effect of the speedup for both scenarios on leptogenesis remains the same. Hence we will only illustrate the result for speedup factor for conformal case as shown in the left plot of Figure \ref{fig:speedup}. 
In this example where speedup happens in the range $10\,{\rm GeV} < T < 10^5\,{\rm GeV}$, it will affect leptogenesis with $M_1$ which falls in the relevant mass range.

Our main results are presented in Figure \ref{fig:ST_bound_M1} in the $K_1 - M_1$ plane for a fixed $m_3 = 0.05$ eV and $v_\nu = 0.1, 1$ GeV (red/thick, blue/thin lines) where outside these closed regime, one cannot obtain sufficient baryon asymmetry.
The solid lines are for $Y_{N_1}(0)=0$ while dotted lines for $Y_{N_1}(0)=Y_{N_1}^{\rm eq}(0)$. 
The horizontal dashed lines are the absolute lower bounds obtained for the respective $v_\nu$ which correspond to having $Y_{N_1}(0)=Y_{N_1}^{\rm eq}(0)$ and no washout.
For $Y_{N_1}(0)=0$ and small $K_1$, due to the speedup in the Hubble expansion, the inverse decay is not efficient in populating $N_1$. Less $N_1$ results in less asymmetry being produced and hence $M_1$ needs to increase correspondingly to enhance the CP violation. As one goes to larger $K_1$, $N_1$ is more efficiently populated and one is allowed to have smaller $M_1$. Crucially, in all cases, $\Delta L = 2$ washout is suppressed sufficiently due to the speedup factor $\xi$ as evidence from the fact that one is able to obtain successful leptogenesis for $M_1$ much below $10^5$ GeV (cf. Figure \ref{fig:GR_bound_M1}). In fact, a smaller speedup in the early times (see Figure \ref{fig:speedup}) allows an efficient washout of an initial `wrong' sign asymmetry (generated during the production of $N_1$) by $\Delta L = 2$ scattering and one ends up enhancing the final asymmetry. Instead of a curse, $\Delta L = 2$ becomes a blessing. Numerically, we found the lowest $M_1$ to be around 350 GeV which corresponds to $v_\nu \sim 0.03$ GeV and $K_1 \sim 3000$. 

As a final remark, notice that the behaviors of $M_1$ lower bounds for $Y_{N_1}(0)=Y_{N_1}^{\rm eq}(0)$ for small $K_1$ are different for the case of $v_\nu = 0.1,1$ GeV. In the small $K_1$ regime, we expect them to approach the absolute lower bounds (the horizontal dashed lines). While this happens for the case of $v_\nu = 1$ GeV, the lower bound actually moves away from the horizontal line for the case of $v_\nu = 0.1$ GeV. The reason is that for $M_1 \sim$ TeV and small $K_1$, the decays happen very late close to the EW sphaleron freezeout temperature $T_{\rm EWSp}$. When we reach this temperature, the baryon asymmetry will be frozen before all $N_1$ can decay, resulting in smaller final asymmetry.

\section{Minimal Supersymmetric Standard Model and right handed neutrinos}
\label{sec:mssm}

In section~\ref{sec:expansion}, we  mentioned that the scale for the enhanced expansion rate can be moved around as a function of the new scale associated with  $D(\varphi)$ term in the disformal case. In the conformal case, an extension of the SM can change the enhancement scale. In this section, we discuss the scale for enhancement in the cases of MSSM and SM with 3 RH neutrinos.

In the left plot of Figure \ref{fig:speedup}, we show the  speedup factor, in the conformal scenario, for  one set of  values for $\varphi$ and $\varphi'$ at a initial temperature of $100$ TeV considering only the SM particle spectrum. If we add three 10 TeV RH neutrinos, the speedup factor and its slope at around 1 TeV is the same as in the SM case, but $\xi$ drops to 1 slightly earlier. 

\begin{figure}[h!]
\centerline{
\begin{tabular}{cc}
\includegraphics[width=0.48\textwidth]{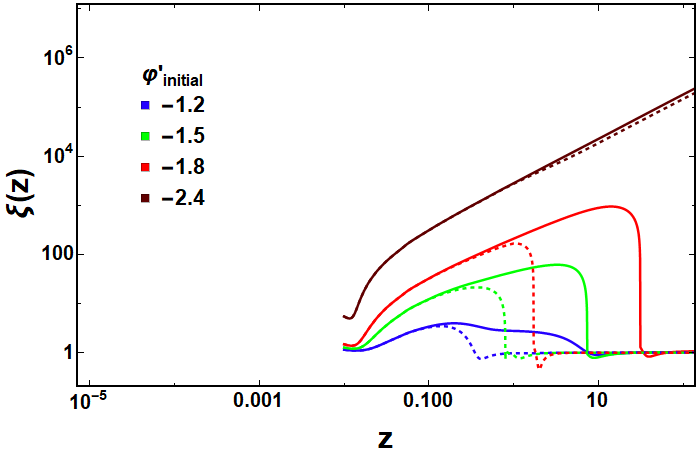}&\includegraphics[width=0.48\textwidth]{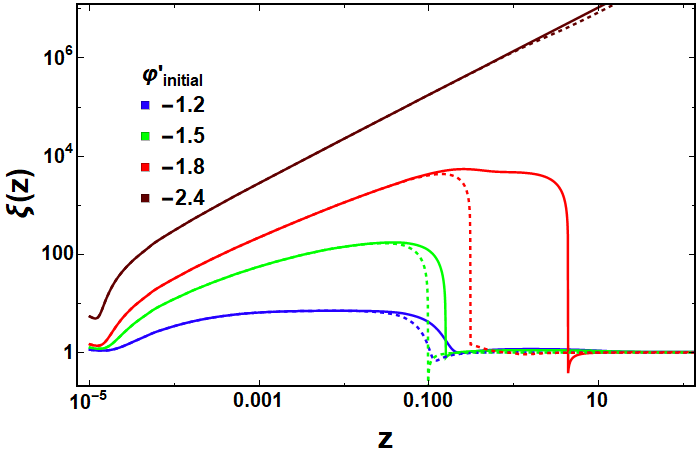}\\
\multicolumn{2}{c}{\includegraphics[width=0.7\textwidth]{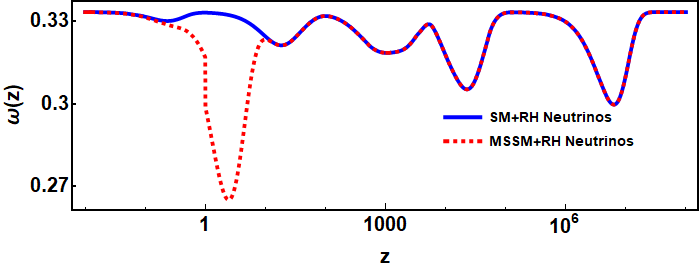}}
\end{tabular}}
\caption{Speedup factor $\xi(z)$ as a function of $z=1\,\rm{TeV}/T$ for SM+3 RH neutrinos (solid lines) and MSSM+3 RH neutrinos (dotted lines) for various values of $\varphi'$. The RH neutrinos mass value is $10\,$ TeV,  and the initial temperatures are $10^5$ GeV (left plot) and $10^8$ GeV (right plot). The bottom figure shows the equation of state parameter $\omega$ for the two cases.} 
\label{fig:speedup1}
\end{figure}

We now add the RH neutrinos  to the spectrum of the SM and MSSM. In Figure \ref{fig:speedup1}, we show the enhancements  for various values of $\varphi'$ and initial temperatures. We add three RH neutrinos at $\sim$ 10 TeV in the particle spectrum along with the SM (solid lines). In the bottom panel of the figure, we show  $\omega$ as a function of $z$ (blue solid line) and we find a new small trough at around 10 TeV due to the new RH neutrinos. The other dips in $\omega$ are due to the SM particles. The expansion rate increases when  $\varphi'$ increases, however if $\varphi'$ is too large, the speedup factor does not get reduced to 1 before the BBN.  As mentioned before, a sudden drop of the enhancement factor to the standard GR value occurs due to the troughs in $\omega$, which create an attractive effective potential when $\omega\neq1/3$. For large initial values of $\varphi'$, the scalar field overcomes this attractive potential  and the  enhancement factor  never reduces to one.

In Figure \ref{fig:speedup1}, left plot, we consider an initial temperature of 100 TeV. The expansion rate can be enhanced by a factor of 100, or more, for temperatures between 100 GeV and 1 TeV. If, however, we increase the initial temperature to $10^5$ TeV or higher, the enhancement scale moves to a higher temperature and  enhancements bigger than 100 can occur for temperatures between 1 TeV and  $10^4$ TeV. At higher temperature, the Hubble friction slows down the scalar field faster, and since the attractive effective potential kicks in early, due to the trough in $\omega$ caused by the RH neutrinos (at around 10 TeV), the enhancement factor drops to one at higher temperature. 

Using dotted lines in Figure \ref{fig:speedup1}, we show the speedup factors due to the MSSM particles, where we keep the SUSY partners of the SM particles at around 1 TeV and, for illustration, three RH neutrinos (and their SUSY partners) at $\sim$ 10 TeV. 
In the panel below the figure, we show  $\omega$ (red dotted line) for MSSM + 3 RH neutrinos and we find a new deep trough in $\omega $ at around 1 TeV due to the SUSY particles (we put all of them together). The other dips are due to the SM particles. Like before, we find that the enhancement and its slope does not change much compared to the SM case but the speedup factor reduces to one for higher temperature.

It is also important to mention, no matter what particle spectrum we consider, that the initial value of the scalar field does not play a relevant roll in the shape and slope of the speedup factor, as long as this value is positive and order one.

We find that an enhancement of the expansion rate with initial temperature at $10^{2-3}$ TeV is most effective in producing a successful leptogenesis, with the enhancement scale around TeV, caused by the SM particle spectrum. Now, the initial temperature is set by the inflation scale. In the case of the MSSM, it is shown that, the thermal leptogenesis constraint from the type I seesaw is $T_R>10^6$ TeV~\cite{Buchmuller:2004nz,Giudice:2003jh}. This bound conflicts with the cosmological gravitino bound for unstable gravitinos. For a gravitino mass closer to a 100 GeV DM mass,  BBN implies stringent upper bound on reheating temperature $\leq O(1)\times 10^2$ TeV~\cite{Kawasaki:2008qe}. Based on our analysis, this low reheating scale is very helpful to have the leptogenesis scale to be around 1 TeV.

\section{Concluding remarks}
\label{sec:conclusions}

Since it is difficult to probe the universe between inflation and the onset of BBN, the evolution of the universe is mostly unconstrained during this period. Origins of DM, baryon abundances etc. are crucially dependent on the evolution history around that time. During this epoch the expansion rate can be different in ST theories compared to the standard cosmology even though the universe is still radiation dominated.   The changes in  the expansion rate  are caused  by conformal and disformal factors in the metric. For an  initial temperature set at  $\sim$ 100 TeV, the conformal modification of the metric can cause an enhancement in the expansion rate, compared to the standard GR case, by more than two orders of magnitude  for temperatures between 100 GeV and a few TeV, due to the SM particles. Although,  the size and shape of the enhancements as a function of time are independent of initial $\varphi$ choices, they depend on initial $\varphi'$ values. The enhancement scale can move to higher temperatures for extensions of the SM and  higher initial temperatures (e.g., $\geq 10^5$ TeV). In the case of a disformal scenario, an enhancement can occur at any scale which is determined by the new scale associated with the $\partial_\mu\phi\partial^\mu\phi$ term in the metric. All these modifications in  expansion rates can cause significant changes in the relic dark matter abundance calculations. In this paper, we focused on the effect of an enhancement of the expansion rate on the scale of leptogenesis.

The scale of leptogenesis in the case of a typical type I seesaw model is very high and is out of reach for the ongoing experimental facilities. However, many models with a much lower leptogenesis scale exist where the RH neutrino masses arise due to new physics around multi-TeV scale. In these models, it is found that (if no resonant enhancement of CP asymmetries is assumed), there exists a lower bound on the scale of leptogenesis which is $\sim$ 10 TeV under the assumption of an initial thermal abundance for RH neutrinos along with no washout. The lower bound increases  in the case of zero initial abundance. This conclusion changes in  ST theories with an enhanced expansion rate which helps the leptogenesis models to be probed in the ongoing experiments.

In the case of an enhanced expansion, the requirement of a larger washout scattering rate demands the scale of leptogenesis to be smaller since the scattering rate is inversely proportional to $M^n$ where the exact value of $n$ ($\geq 0$) depends on the details of the initial and final state particle properties. We used a toy model of leptogenesis to manifest the lowering of the leptogenesis scale due to an enhanced expansion  rate.  In this model the RH neutrinos do not couple to the SM Higgs, instead they  couple to a new Higgs. We found that the scale of leptogenesis can be lowered down to $\sim$ TeV for both zero and thermal initial abundances for the RH neutrinos for a wide range of model parameter space  which allows these models to be probed at the ongoing experimental facilities. In some parameter space of the  model, we showed that an enhancement of the expansion rate can lower the leptogenesis scale down to $\sim 400$ GeV. The existence of an enhanced expansion rate between 100 GeV to a few TeV  due to the SM particle spectrum (plus the RH neutrinos) in the case of a conformal modification of the metric is crucial to lower the scale of leptogenesis. If an enhancement happens at a higher scale, the scale  of leptogenesis is not lowered and an enhancement at a smaller scale is also not helpful in lowering the leptogenesis scale with the correct amount of asymmetry since the EW sphaleron freezeout occurs at around 130 GeV. All of our findings for this model should apply to any leptogenesis model.

\section*{Acknowledgments}
B. D. and  E. J. are supported in part by the DOE grant DE-SC0010813. B. D. and E. N. acknowledge the kind invitation of D. A. Restrepo to visit the Universidad de Antioquia in Colombia where this work was initiated.  The work of E. N. is supported in part by the
INFN ``Iniziativa Specifica'' Theoretical Astroparticle Physics
(TAsP-LNF).
C. S. F. was supported by the S\~{a}o Paulo Research Foundation (FAPESP) grants 2012/10995-7 \& 2013/13689-7 and 
is currently supported by the Brazilian National Council for Scientific and Technological Development (CNPq) grant 420612/2017-3.

\bibliography{CosmoLepto}

\end{document}